\documentclass[10pt]{IEEEtran}

\usepackage{amsmath}
\usepackage{bm}
\usepackage{amssymb}
\usepackage{latexsym}
\usepackage{multirow}
\usepackage{epsfig}
\usepackage{graphics}
\usepackage{mathrsfs}
\usepackage{algorithm}
\usepackage{balance}
\usepackage{algpseudocode}
\usepackage{subcaption}

\usepackage{color}
\usepackage{graphicx}
\usepackage{algcompatible}
\usepackage{array}
\usepackage{amsfonts}
\usepackage{mdwmath}
\usepackage{mdwtab}
\usepackage{eqparbox}
\usepackage{epstopdf}

{}
{}
{}

\begin{document}

\title{Exploring Communication Technologies, Standards, and Challenges in Electrified Vehicle Charging}

\author{(\textit{Invited Paper}) \\ \IEEEauthorblockN{Xiang Ma\IEEEauthorrefmark{1}, Yuan Zhou\IEEEauthorrefmark{1}, Hanwen Zhang\IEEEauthorrefmark{2}, Qun Wang\IEEEauthorrefmark{3}, Haijian Sun\IEEEauthorrefmark{2}, Hongjie Wang\IEEEauthorrefmark{1}, Rose Qingyang Hu\IEEEauthorrefmark{1}\IEEEauthorrefmark{4} }
\IEEEauthorblockA{
\IEEEauthorrefmark{1}Department of Electrical and Computer Engineering, Utah State University, Logan, UT \\
\IEEEauthorrefmark{2}School of Electrical and Computer Engineering, University of Georgia, Athens, GA\\
\IEEEauthorrefmark{3}Department of Computer Science, San Francisco State University, San Francisco, CA\\
\IEEEauthorrefmark{4}Email: rose.hu@usu.edu}

}



\maketitle

\begin{abstract}
As public awareness of environmental protection continues to grow, the trend of integrating more electric vehicles (EVs) into the transportation sector is rising. Unlike conventional internal combustion engine (ICE) vehicles, EVs can minimize carbon emissions and potentially achieve autonomous driving. However, several obstacles hinder the widespread adoption of EVs, such as their constrained driving range and the extended time required for charging. One alternative solution to address these challenges is implementing dynamic wireless power transfer (DWPT), charging EVs in motion on the road. Moreover, charging stations with static wireless power transfer (SWPT) infrastructure can replace existing gas stations, enabling users to charge EVs in parking lots or at home. This paper surveys the communication infrastructure for static and dynamic wireless charging in electric vehicles. It encompasses all communication aspects involved in the wireless charging process. The architecture and communication requirements for static and dynamic wireless charging are presented separately. Additionally, a comprehensive comparison of existing communication standards is provided. The communication with the grid is also explored in detail. The survey gives attention to security and privacy issues arising during communications. In summary, the paper addresses the challenges and outlines upcoming trends in communication for EV wireless charging.

\end{abstract}

\section{Introduction}\label{sec1}
Paris Climate Agreement aims to achieve carbon neutrality by the year 2050 \cite{paris}. To achieve this goal, a crucial element is the transition from conventional gasoline-consuming vehicles to electric vehicles (EVs) in the transportation field. EVs utilize batteries, ultracapacitors, and fuel cells as energy sources rather than fossil fuels, resulting in clean emissions and a minimal carbon footprint. According to a report \cite{iea_report} by the International Energy Agency (IEA), the number of EVs will expand to almost 350 million by 2030 worldwide, which brings both opportunities and challenges for the automotive industry. One of the primary challenges is the low energy density of the batteries currently employed in mainstream EVs, which require frequent charging of EVs. Additionally, the substantial costs of deploying EV charging infrastructures impede investment progress. Consequently, EVs are experiencing increased concerns about the driving range.  

In the traditional method of EV charging, vehicles are replenished through a cable, which necessitates a physical connection between the EVs and the charging infrastructure. This direct connection poses several practical challenges, including the demand for large charging areas and the potential traffic congestion \cite{lee2020exploring}. An alternative solution for cable plug-in charging is wireless power transfer (WPT), which charges EVs using air instead of conventional cables. It can be categorized as near-field and far-field WPT \cite{zhang2018wireless}. The far-field WPT can be realized using acoustic, optical, or microwave as the carrier, although it is usually less efficient. The near-field WPT uses the inductive/capacitive coupling effect of electromagnetic fields, including inductive coupling, resonant inductive coupling, and capacitive coupling. Inductive coupling and resonant inductive coupling, called inductive power transfer (IPT), utilize the magnetic field, while capacitive coupling applies the electric field to deliver the energy. Capacitive coupling requires high voltages on the electrodes and can be hazardous, so it is not practically used in high-power transfer. The primary EV wireless charging applies IPT to deliver energy. The principle of IPT is to construct a wireless power transfer system with coupling devices. The power transmitter is connected to a power source, while the receiver is linked to the intended load. When the coupling devices on both the transmitter and receiver sides are close enough, the power is transferred through magnetic flux \cite{covic2013inductive}. In WPT systems designed for EVs, the configuration typically includes one charging coil on the board while another is mounted within roads or mobile energy sources.

\begin{figure*}[!th] 
    \centering
        \includegraphics[width=\linewidth]{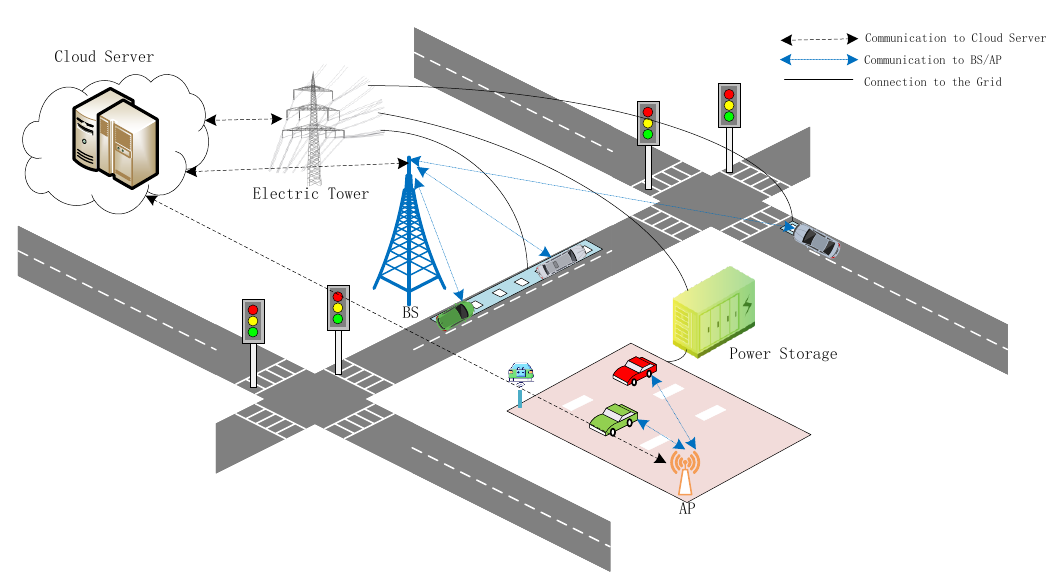}
        \caption{Communication for Electric Vehicle Wireless Power Transfer}
        \label{fig:system_diagram}
        
\end{figure*}

In the WPT domain for charging EVs, there are three operational modes: 1) static WPT (SWPT), 2) dynamic WPT (DWPT), and 3) quasi-dynamic (QDWPT) \cite{mahesh2021inductive}. SWPT occurs when EVs remain static at parking lots, charging stations, and other areas. Similar to plug-in charging, there are designated spots for each EV. Conversely, DWPT allows EVs to charge while moving, which avoids long waiting times for charging. QDWPT, on the other hand, is tailored for scenarios that are either static or moving at low speeds for short periods. The paradigm suits EVs halting at bus stops, taxi stands, or traffic signals. 

In addition to transferring energy from sources to EVs, the emergence of the vehicle-to-grid (V2G) could offer a solution to grid-related challenges, which is achieved by supplying the grid with electricity stored in EVs, thereby providing resources for managing energy demands and enhancing grid stability \cite{8723851}.

Although WPT provides a promising solution for charging EVs without the restrictions of a physical connection, it encounters several non-trivial issues. Among these, a key concern is the requirement for precise real-time control for achieving continuously changing coupling between the receiver coil and the transmitter coil, which is a critical factor that influences the efficiency of wireless charging systems \cite{7559721}-\cite{7925821}. Alongside control-related challenges, privacy and security issues need to be considered due to the reliance on wireless communication. The broadcast nature of wireless signals increases the vulnerability of WPT charging systems, which raises security concerns, including the vulnerability to various attacks like eavesdropping, man-in-the-middle, replay, impersonation, and denial-of-service (DoS). These attacks threaten user privacy and the integrity of financial and operational data. Therefore, ensuring robust authentication and securing communication processes are critical to safeguard against these vulnerabilities. Load balancing in the power grid is another significant challenge posed by adopting WPT. The uncoordinated charging of a large number of EVs can lead to increased power losses and voltage fluctuations. Such issues could potentially overload the power grid, underscoring the need for careful management of grid load to maintain stability \cite{6847105}. Lastly, developing a fair and secure trading framework is necessary to maximize the benefits for buyers and sellers in a dynamic charging system \cite{AUJLA2019169}. While existing wireless communication methodologies have provided practical solutions for vehicular communication, the dynamic nature of EVs, coupled with the unique communication characteristics for WPT charging, presents novel complexities. The high mobility of EVs not only introduces new challenges on fast-changing communication channels, resource management, and handover but also exacerbates the issues above. Moreover, the short period during which the effectiveness of power transfer is acceptable, and power can be transferred imposes higher requirements for wireless communication management and coordination. Consequently, there is a critical need for mobility awareness in dynamic charging systems \cite{wevj12030092}.

To effectively address these challenges and improve the overall efficiency of the charging system, it is crucial to establish highly reliable, low-latency, and high-capacity communication links within various parts of the WPT systems. Different types of network connections among different entities can be utilized to satisfy the communication demands for WPT services: On-board service requires the communication connection between vehicles and infrastructure (such as base station and roadside units), named Vehicle-to-Infrastructure (V2I). On the other hand, the second group of services with connectivity requirements is vehicle-to-vehicle (V2V) solutions. For this type of service, the most extended technologies are based on ad hoc networks applied to the vehicle field (VANET or Vehicular Ad hoc NETworks). In some services for charging, the messages need to be delivered from one infrastructure to another. This kind of connection is called Infrastructure-to-Infrastructure (I2I) networks. Integrating these communication connections into the wireless charging system is essential to ensure seamless power transfer, real-time data exchange, and operational coordination between the entities involved. By establishing the management and coordination based on these communication links, the WPT system can better adapt to the dynamic demands of EVs, thereby improving the efficiency and reliability of power transfer processes \cite{wevj12030092}. For security concerns, implementation of authentication, encryption, position verification, digital certificates, and cryptographic protocols is required to ensure the safety and reliability of WPT systems. 

Various related protocols and communication standards are proposed in V2X to support vehicular communication, notably the Dedicated Short-Range Communication (DSRC) and the Cellular-based Vehicular Network (C-V2X). DSRC, which encompasses IEEE 802.11p and IEEE 1609.1.4, is mainly used for V2I communication, whereas C-V2X, proposed by 3GPP, focuses more on V2V and V2P communications. In addition, slower speed or short-range communication standards such as ZigBee, LoRa, Bluetooth, and WiFi are used in the V2X field, particularly in static EV wireless charging. 

In the last decade, communication technologies related to EV charging have attracted significant interest from both the industry and academia. In \cite{wevj12030092}, the existing communication technologies for coordinating and managing EV charging are investigated. In addition, the physical layer security strategies for EVs are discussed. However, it fails to analyze the characteristics and requirements of the communication for different WPT systems in detail. The authors of \cite{7056176} investigate the vital characteristics of communication systems in the context of DWPT communication protocols and system architecture. Nevertheless, the coverage of the paper is limited. There has been a lack of comprehensive review papers addressing communication techniques in the context of electric vehicle wireless charging. In this survey, we focus on the communication involved in EV wireless charging, including EV detection, charge initiation, charge termination, and payment. We explore the characteristics and requirements of communication under SWPT and DWPT. The existing communication standards will also be compared regarding range, data rate, latency, and other aspects. Then, the communication from EVs to the grid is investigated to study the effects of WPT on the grid. Finally, the communication security and privacy issues are studied. 


\section{Communication for static wireless charging}\label{sec2}
This section will first discuss current wired charging technologies and then introduce wireless charging, specifically focusing on SWPT. Following that, we will conduct a comprehensive survey of the communication aspects associated with SWPT, including topics such as vehicle identification, charging initiation, charging status communication, and more. Lastly, we will discuss the communication requirements for SWPT.

\subsection{Wired and wireless charging}
\begin{figure}[!th]
\centering
\includegraphics[width=\linewidth]{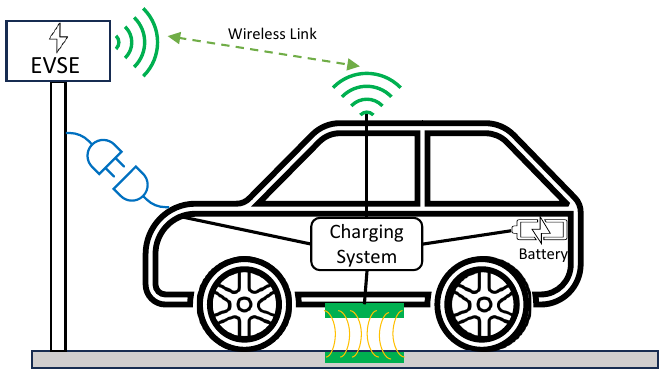}
\caption{Wired and Wireless Charging Diagram}
\label{fig:swpt_diagram}
\end{figure}

The mainstream charging method for EVs is plug-in wired charging. One end of the electric cord is connected to the power source, and the other can be plugged into EVs. As shown in Fig. \ref{fig:swpt_diagram}, the blue cord can charge the battery of the EV through the charging system. For wireless charging, electromagnetic fields are used as power transmission media. The mechanism for power transmission is either capacitive coupling or inductive coupling. Since there is no wire connection between EVs and EVSE, there should be a standalone wireless communication link to control the charging process, as shown in Fig. \ref{fig:swpt_diagram} (green part).

\subsubsection{Wired charging}
From the perspective of electric current, wired charging can be alternating current (AC) or direct current (DC). The typical standard SAE J1772 is used for AC charging, while the combined charging system (CCS), CHAdeMO, and Tesla's supercharging are defined for DC charging. Each standard has multiple pins in the connector for power supply and charging communication. For example, in SAE J1772, there are five pins in total. Two of the pins are for the power supply. Another is the ground pin to protect the EVs. And the other two are for charging communication. One communication pin is called the proximity pilot (PP) for pre-insertion signaling. This allows EVs to ascertain their connection to the EV supply equipment (EVSE).
 Another communication pin is the control pilot for post-insertion signaling. It serves the purpose of negotiating the charging amount between the EV and EVSE, initiating the charging process, and conveying additional information.

\subsubsection{Wireless charging}
In wireless charging, the mainstream technique is inductive coupling, which requires two charging pads/coils for power transmission. The primary charging pad is mounted on the ground for power transmission, and the secondary charging pad is attached to the vehicle for power reception. According to different implementations, the secondary pad can be attached underneath the vehicle \cite{panchal2018review} or as the coil array format installed in-wheel \cite{panchal2017static}. Magnetic fluxes carry the power from the primary to the secondary pad. Therefore, the alignment of the primary and secondary pads can significantly affect the charging efficiency. 

There is no wire connection needed in wireless charging. The communication between EVSE and EVs should also be wireless. The specific features of wireless charging bring challenges, such as pad alignment and security/privacy issues in wireless communication, but also create new opportunities to reduce manual intervention with plugging and unplugging the EVs to EVSE, hence reducing electrical safety concerns. 

\subsubsection{SWPT} In SWPT mode, the EV is parked in a fixed place and remains static during charging. Each EV has a standalone charging spot like in wired charging. It can happen at a charging station, home garage, or a designated shopping mall parking lot. The driver can charge the EV when parking it. There are multiple advantages of SWPT among wired charging. First, with engineering progress, SWPT can achieve $95.8\%$ power conversion efficiency at $50$ kW \cite{bosshard2016multi}, comparable with wired charging.  Next, Due to its wire-free nature, SWPT can be used in various weather conditions, including rain, snow, and ice \cite{fisher2014electric} without worrying about electrical damage. 

\subsection{Communication involved} \label{sec2_2}
To enable SWPT, the electric circuit of the EVSE needs to be well-built. The power electronics of SWPT have already been well investigated in \cite{zhang2015loosely, ahmad2017comprehensive, kan2018integrated}. The coil design, AC/DC and DC/AC converter design, and the compensation network are all essential components for SWPT. However, charging control is important but has yet to be well studied. This section will thoroughly survey the communication between EVSE and EVs during the SWPT process. 

The initial stage of SWPT involves coil detection, which signals the EVSE that the EV is positioned above the primary pad. So the EVSE can prepare the power for charging. Subsequently, to enhance charging efficiency, precise coil alignment is necessary. Several methods can be applied to address the misalignment issue. First, multiple primary coils \cite{liu2007equivalent} or multiple secondary coils \cite{budhia2011development} are used to provide more alternative coils for power transfer. However, this method adds complexity and cost to the whole system. Another method is to move the primary coil with a moving mechanical device \cite{cook2019wireless}. This provides convenience for EV drivers and does not require them to park in specific areas. However, mechanical movable devices add additional energy costs. Another possible solution is an automatic guidance system to help drivers align the vehicle above the primary pad. Cameras, ultrasonic sensors, radio frequency identification (RFID) sensors \cite{SunRFID}, or sensing coils are different options for automatic alignment. Recently, a self-aligned method has been proposed \cite{li2023robust} for automatic-guided vehicles. Wireless communication provides feedback to the primary controller on the side and helps the vehicle adjust its position.

Before charging initiation, authentication should be performed. In a wired charging scenario, such as Tesla's supercharging, the vehicle identification number (VIN) needs to be sent from the vehicle to the EVSE. Only the authorized vehicle is allowed to use the EVSE. The raw identification information should be hidden during transmission for the wireless charging scenario to avoid identity leaks. Then, charging power and charging price negotiation are performed. Usually, a high charging power comes with a high unit price. The vehicle transmits its current battery information to the EVSE, enabling the calculation of an estimated time for charging completion.

When the charging preparation work is done, the primary pad is powered on by the EVSE, and the energy is transferred from the EVSE to the vehicle. And the battery status of the EV is sent to EVSE. When charging is completed, an EV termination signal may be sent to avoid overcharging.

After charging, the payment can also processed via wireless communication in public charging stations. Since the vehicle identity is known by the EVSE, the operator of the public charging station can charge from the account associated with the vehicle ID. This also reduces human intervention and makes the charging process more convenient.

\begin{figure}[!th]
\centering
\includegraphics[width=1\linewidth]{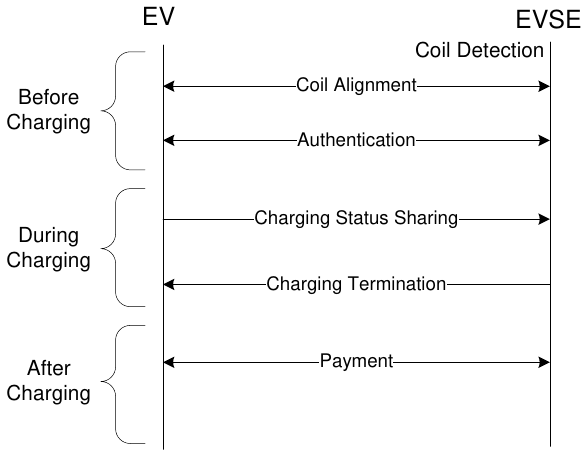}
\caption{SWPT Communication}
\label{fig:swpt_comm}
\end{figure}

\subsection{Communication requirements}

As mentioned in section \ref{sec2_2}, the vehicle identification, charging authentication, and battery status information transmission involved in SWPT require a reliable, high-throughput, and safe communication link. The communication performance requirements for SWPT can be summarized as follows:
 \begin{itemize}
    \item \textbf{Latency:} According to the US Department of Energy, the end-to-end communication latency for static EV charging should be in the second level \cite{us_department_energy}. This should be comparable with wired charging to attract users. In large public wireless charging stations, wireless channel resource allocation should be considered to avoid congestion and delays. 

    \item \textbf{Throughput:} For the charging control communication, the throughput of 1 Mbps can satisfy the requirements \cite{us_department_energy}. It requires less bandwidth than other applications, such as video transmission. 

    \item \textbf{Reliability:} Reliability is significant for communication during the charging process. The EVSE needs to monitor the status of the battery and terminate the charging when it finishes or whenever necessary. It provides safety protections. 

    \item \textbf{Security and privacy:} The communication between EVSE and EV should be secure. The privacy-sensitive data such as EV location, EV ID, and payment should be encrypted. The communication should consider eavesdropping, jamming, and other physical layer attacks.
\end{itemize}

\subsection{Current related standards}
Several existing standards describe SWPT in terms of communication. SAE J2954 \cite{sae_j2954} defines WPT for light-duty EVs and alignment methodologies. It briefly describes the communication process between the EV and EVSE. EV and EVSE should communicate with each other for positioning and coil alignment. And the charging status is constantly shared during the charging process. However, the communication details are not defined. 

SAE J2847-6 \cite{sae_j2847_6} discusses the communication requirements between EV and WEVSE for SWPT. The communication messages and procedures are defined. Different message types for SWPT communication are specified. However, the communication is not specified as wireless communication, so the design lacks wireless communication thought.

ISO 15118-8 \cite{iso_15118_8} employs IEEE 802.11n wireless communication technology for communication between EV and EVSE. The EVSE is required to be configured as an access point (AP), and EVs can connect the AP through WLAN protocols. The communication range, channel, and timing can be inherited using the existing wireless communication technology. 

Among all the standards mentioned above, the communication for SWPT is still under exploration. 

\section{Communication for dynamic wireless charging}\label{sec3} 

Given the advantages and limitations inherent to various dynamic wireless charging methodologies, it becomes imperative to incorporate these systems into infrastructure and EVs to meet the growing EV charging demands in different scenarios. Wireless communication-enabled energy management, monitoring, and control are essential to prevent disruptions and overloads in charging processes. Consequently, establishing an interconnected network among EVs and infrastructures \cite{wevj12030092} is critical. In this section, existing DWPT and QDWPT methods are introduced first. Then, the communication involved in WPT is analyzed to capture the characteristics of the communication, followed by a discussion on the associated communication requirements.

\subsection{DWPT and QWPT}
Despite the merits of SWPT in providing power transfer without physical connection, it fails to address the issues of frequent charging requirements and the need for large battery capacities in vehicles. In contrast, utilizing DWPT while the vehicle is in motion could offer an infinite driving range without requiring large battery capacities \cite{101049}. On-road EV charging can be categorized into three types: G2V-based QDWPT, G2V-based DWPT, and V2V-based DWPT \cite{wevj12030092}. In G2V-based paradigms, energy transfer occurs from a primary coil embedded under the surface of the road, powered by the electrical grid \cite{8745768}. Alternatively, V2V represents a paradigm where power is transmitted from coils mounted on mobile energy disseminators (MEDs) to EVs \cite{9383634}.

\subsubsection{QDWPT}

QDWPT is a hybrid of dynamic and static charging where the EV is charged during transient stops or slowly moving at intersections and traffic signals. The power transfer occurs through the interaction between a charging coil on the vehicle and coils embedded under the road surface, where vehicles tend to move slowly. A significant benefit of QDWPT is that it does not require vehicles to park at charging stations or parking lots for charging. Moreover, the slow movement of vehicles at intersections creates a relatively static charging environment. As a result, QDWPT can leverage the benefits of both SWPT and DWPT.

Compared to SWPT, QDWPT mitigates the need for frequent charging intervals and reduces the reliance on large-capacity batteries in EVs \cite{1010491}. However, the involved dynamic processes introduce extra complexity into the electromagnetic environment, which needs to be supported by more sophisticated infrastructures to ensure efficient and safe operation. 

\subsubsection{DWPT}
Although quasi-dynamic charging provides a more flexible method for on-road charging, its application is limited to intersections. Based on the entities involved in the charging process, there are two basic methods of DWPT for charging EVs in motion: G2V DWPT and V2V DWPT.

In G2V DWPT, EVs are charged while in motion on charging lanes that have the coils/pads embedded under the road surface, which can be implemented in a broader range of scenarios, including highway and urban areas \cite{GARCIAVAZQUEZ201742}. The G2V DWPT system shares components similar to the G2V QDWPT system. Nevertheless, there are several additional challenges in the deployment of G2V DWPT systems, particularly the coil alignment \cite{8217218}, authentication \cite{s6}, resource allocation, and handover with high mobility and density \cite{wevj12030092}.

Unlike G2V-based DWPT methods, V2V DWPT systems do not rely on costly infrastructures. This characteristic facilitates the application of the V2V DWPT paradigm in particular scenarios like wilderness and rural areas. In V2V DWPT, vehicles such as buses or trucks serve as mobile energy sources for EVs requiring charging. These MEDs utilize IPT technology to replenish the batteries of EVs that require energy \cite{moschoyiannis2015dynamic}. The DWPT charging process is realized by integrating coils on both MEDs and EVs, with onboard controllers (OBCs) equipped to close the control loop on both sides. However, challenges such as routing, synchronization, and health and safety concerns need to be addressed to achieve fast and effective V2V DPWT \cite{6850600}.

\begin{figure*}[!th] 
    \centering
    
        \includegraphics[width=\linewidth]{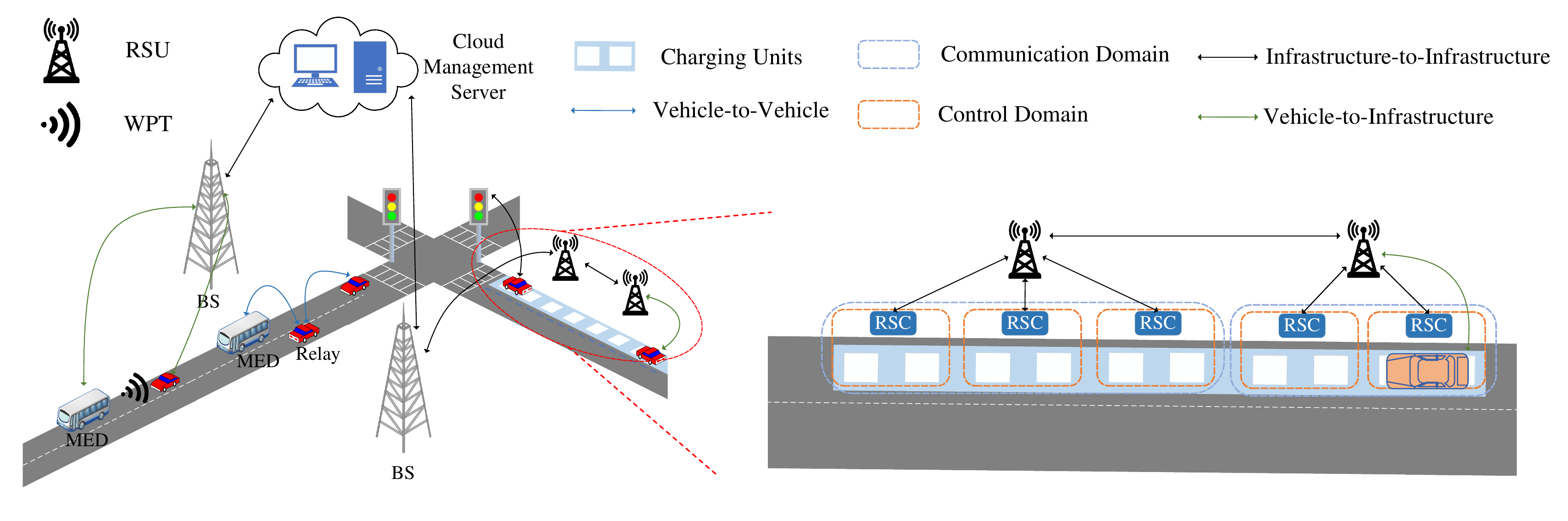}
        \caption{DWPT Communication}
        \label{fig:dwpt_comm}
        
\vspace{-0.55cm}
\end{figure*}

\subsection{Communication involved}

G2V QDWPT, G2V DWPT, and V2V DWPT are illustrated in Fig. \ref{fig:dwpt_comm}, where the wireless communication links play a critical role in integrating charging infrastructures with EVs for dynamic wireless charging. 

In G2V QDWPT and DWPT, primary coils are embedded in designated areas named charging lanes. The charging infrastructures may belong to different equipment providers within these charging lanes. To manage the charging process effectively, the concept of control domain is introduced, defined as the area controlled by a roadside controller (RSC). Collecting control domains belonging to the same equipment provider constitutes a specific charging system (VSCS). One roadside unit (RSU) can cover several control domains owned by the same or different VSCS. 

The messages transmitted vary based on the particular design of the charging management frameworks and the specific charging scenarios. Generally, two categories of data streams are exchanged in the majority of G2V DWPT and QWPT systems according to the range of communication: Messages require wide-area communication, such as authentication and trading messages, which are transmitted through the backbone network; Short-range communication messages that include information like charger location, status, which are necessary for the controlling of the power transfer process \cite{6994754}.

G2V QDWPT and G2V DWPT share a similar charging process, divided into two stages: Discovery and charging control (including coil alignment and power control) \cite{8644359}. Once EVs enter the communication range of the communication supply infrastructures for charging (i.e., base stations or roadside units), vehicles should access the infrastructures to enable registration and authentication through V2I communication and I2I communication. The dynamic nature of G2V DWPT introduces additional complexities in channel access and authentication compared to conventional communication scenarios, particularly concerning key management \cite{s6}. Furthermore, the authentication and trading associated with multiple charging lanes, each owned by different VSCSs, requires the establishment of either centralized or distributed management and coordination frameworks to ensure cohesive and efficient charging processing. A significant challenge is the necessity for low-delay and high-reliability handovers crossing different communication domains and VSCSs due to the high mobility of EVs. Moreover, the dense connectivity of EVs creates a resource-constrained environment, potentially degrading the communication performance and further causing the failure of the charging system. Consequently, effective and fast resource allocation strategies, including power slitting, channel access, and RSU allocation, should be performed to establish reliable communication and power links among EVs and charging infrastructures \cite{9134383}. In the second stage, a control loop between primary and secondary coils should be established for state monitoring, charging power control, and coil alignment \cite{6994754}. Typically used primary- and dual-side closed-loop controls in this stage require a communication link between the EV and the coil to transmit feedback data. The feedback data from EVs leads to more accurate results than estimations and adjustments made on the primary side. Alternatively, secondary-side control can avoid the need for wireless communication, thereby eliminating the concerns of communication latency. Wireless communication can be realized either through a communication link from the EV to the RSU and then to the coil or through a directed communication channel between the EV and the coil.


Based on the cooperative framework proposed in \cite{6850600}, the charging process in V2V DWPT can be divided into four stages: Discovery, negotiation, routing, and charging. During the discovery phase, EVs actively search for MEDs within the traveling range to schedule charging appointments while traveling. MEDs periodically broadcast cooperative awareness messages (CAM), which propagate through the VANETs or cellular network, to announce their location and available charging slots. When an EV requires charging from a MED, it first verifies whether the MED's route aligns with its current route. Then, the EV evaluates if the charging carrying capacity of the MED is sufficient to meet its energy needs in the negotiation phase, where the MED and EV exchange information to determine the energy transfer arrangement. The EV requests a charging slot by a CAM that specifies the minimum required charging duration. Decides to designate the MED as the selected wireless energy transfer station. Then, the EV reserves a charging spot. During these two stages, the communication involved has the nature of tolerance for delays, thus allowing for opportunistic V2V communication. However, unlike static charging scenarios where the location of the destination of the message is fixed, MEDs are in motion, introducing extra dynamics into the routing of messages. On the other hand, authentication and trading mechanisms need to be established to address issues such as forging and false messages, privacy leakage, and maximizing the benefits of both the energy provider and receiver. The wireless communication for the reservations of static charging has been fully explored. In \cite{8734897}, the charging reservation via opportunistic and deterministic transmissions is compared, which shows that opportunistic V2V communication can reduce the peak load on cellular networks. Although deterministic communication can offer better reliability and lower delay, the delay tolerance nature of the reservation information makes feasible and cost-efficient opportunistic V2V communication a viable and cost-effective option. Based on VANET infrastructure, the authors of \cite{komala2023vanet} proposed a distributed P2P EV authentication and trading solution where security, reliability, and availability are considered. However, the high mobility of both the EVs and the MEDs poses challenges in designing the trust model for VANETs. It requires further exploration into the wireless communication-enabled cooperative and management mechanisms. In the charging stage, EVs are either in front of or behind the MED for the agreed duration to recharge. V2V DWPT faces challenges, such as short contact periods, rapidly moving nodes, simultaneous charging of multiple nodes, and interference \cite{6850600}. These operational challenges require frequent information exchange between MEDs and charging EVs during the charging process. In addition, EV booking charging services may need to alter their routes to follow or lead a bus, potentially causing traffic flow density fluctuations and the formation of vehicle clusters around MEDs \cite{8402042}.

Autonomous driving makes it possible to precisely control the behaviors of EVs, which enables the optimal design of the DWPT system for eco-driving electric vehicles at urban intersections and the influence of driving behaviors on charging efficiency \cite{8644359}. In \cite{1010491}, V2I communication is utilized by the infrastructure to send traffic timing information to connected autonomous vehicles (CAV). When it enters the range of V2I communication and approaches the intersection, the CAV changes its driving behaviors to minimize the number of cars stopped at the signal stop line and avoid abrupt speed changes. If a CAV follows a CAV, the front CAV informs the intended speed of the followed CAV. The results demonstrate that behavior control reduces energy consumption and increases the average amount of transferred energy. A key challenge for V2V DWPT is keeping the optimal distance between MEDs and EVs for effective charging while ensuring safety. Hence, coordination based on precisely controlled autonomous driving is promising for solving this issue, which has not been explored.



\subsection{Communication requirements}
Existing communication infrastructure, like cellular networks, can be utilized for WPT communication. The authentication and billing data exchange requires a bidirectional communication link between EVs and POs, which can be realized via cellular network base stations. The sizeable cellular network coverage can avoid the handover problem by covering the whole charging lane. Although cellular networks are highly dependent on service providers, they fit most of the scenarios of G2V DWPT. In addition, EVs and POs can communicate through power-line communication (PLC) to provide high-speed broadband communications \cite{s6}. On the other hand, V2V DWPT systems exhibit a preference for utilizing VANETs over conventional cellular networks. The preference is attributed to VANETs' advantages, including reduced communication costs, no energy limitations, superior scalability, and a high degree of self-organization \cite{8913525}. However, in certain specific scenarios where traffic density is low, VANETs may necessitate support from cellular or satellite networks to establish a backbone network \cite{9523788}.

The charging modes and the type of data streams involved mainly determine communication requirements for DWPT. Control-related messages are critical for ensuring an effective charging process and, thus, should be delivered with low delay and high reliability. While other types of data do not have restrictive constraints for the latency and reliability \cite{7056176}.

For G2V DWPT, the system faces the most significant mobility challenges, particularly in managing the handover between RSUs and efficiently handling resource allocation for EVs that are dynamically distributed across various geographical locations. The protocol or networking topology should consider the time gap between the handover crossing two communication domains. 

In a quasi-dynamic WPT scenario, the charging systems are deployed at intersections in cities with dense connections. Since charging occurs during transient stops or slow movement at intersections, the demands for mobility and low latency are comparatively lenient. However, the messages of control commands still require low latency and high-reliability communication.

V2V DWPT has different requirements for reservation messages and control commands. During the discovery, negotiation, and routing stage, the system allows packet loss and delays using scheduled information and requires a wide range of communication methods. In contrast, the necessity for real-time control information becomes paramount, demanding a continuous and reliable exchange of data in the charging stage. This is crucial to ensure efficient wireless charging and maintain safe distances between MEDs and EVs, particularly in autonomous driving scenarios. In addition, it is crucial to identify the significance and timeliness of different data streams. However, most of the current existing communication systems lack content-centric intelligence, especially in the lower protocol layers.

For all the scenarios and data types, dynamic charging systems need to preserve the location privacy and billing and authentication secure during the communication between the EV and other parts of the dynamic charging systems and involved institutions \cite{wevj12030092}. Due to the high mobility, the authentication in DWPT is more challenging than that in SWPT. Capacity is another important communication feature. Although most data streams have small packet sizes, DWPT systems require large communication capacity to serve all the data streams for a single vehicle \cite{us_department_energy}.



\section{Communication standards for EV charging}\label{sec4} 
To integrate EV charging with charging management systems, vehicle-to-everything (V2X) communication presents a potential solution. In this section, we will discuss the various standards associated with V2X and the communication requirements for two charging scenarios: SWPT and DWPT. 
\subsection{Comparison} \label{sec4_1} 
Over the past few years, the ubiquitous communication standards in the V2X field are the DSRC \cite{DSRC_intro} and the C-V2X \cite{CV2X_intro}. The DSRC standard includes IEEE 802.11p and IEEE 1609.1.4 and involves the applications in V2X, such as resource management and wireless access \cite{DSRC_overview}. On the other hand, the 3GPP proposed the C-V2X \cite{Standard_timeline}, a standard that shares similarities with DSRC but emphasizes different areas. C-V2X involves more communication on V2V and vehicle-to-pedestrian (V2P). In addition to these primary V2X communication techniques, slow-speed or short-range communication standards, including ZigBee, LoRa, Bluetooth, and wireless fidelity (WiFi), are also utilized in the V2X field, particularly in EV wireless charging. Despite the limitations of their low data rate, relatively high delay, or short range, these standards can still share the communication load with high-speed communications or be applied to some long-term applications due to their unique property.

\begin{table*}[h!]
\centering
\caption{Comparison of Different Communication Method}
\label{tab:Comm_Comparason}
\begin{tabular}{c|c|c|c|c|c|c|c}
    \hline
    Method& Data rate (bps) & Delay  & Spectrum (GHz) & Cost &Range& Organization & Standard\\ \hline
     BLE      & Low             & Low &2.402-2.48 & Low & Short& SIG&IEEE802.15.1\\ 
     Bluetooth& Mid             & Low &2.402-2.48 & Relatively Low& Short& SIG&IEEE802.15.1\\ 
     WIFI 6/6E& Very High  & Very Low &2.4-6 & Mid & Mid&IEEE &IEEE802.11\\ 
     LoRa     & Low             & High & 0.902-0.928 & Low& Extremely Long & LoRa Alliance& ITU-T Y.4480\\ 
     ZigBee   & Low             & High &2.4, 0.915 ,0.868 & Low& Short&  CSA&IEEE802.15.4\\
     DSRC     & High            & Low &5.850-5.925 & High&  Mid& IEEE&IEEE802.11p\\
     C-V2X    & High& Low &0.450-7.125 & High& All Cover & 3GPP&3GPPRelease14\\
     NR-V2X   & Very High  & Very Low &0.450-7.125, 28-52 & Very High& All Cover& 3GPP&3GPPRelease16\\
     \hline
    
\end{tabular}
\end{table*}
\subsubsection{DSRC and C-V2X}
Due to the utility of DSRC and C-V2X, they are widely used communication standards in V2X. In this subsection, we introduce DSRC and C-V2X separately. First, a concise overview is provided for each standard. Subsequently, state-of-the-art research for DSRC and C-V2X surrounding EV charging has been given. Finally, we indicate some disadvantages in DSRC and C-V2X in EV charging communication.

\textbf{DSRC}:
DSRC is firstly designed to provide communication among vehicles and infrastructures \cite{Standard_timeline}.  While in Europe, it is called intelligent transportation systems generation 5 (ITS-G5). This system has two sorts of devices, OBU and RSU, serving V2V and V2I, respectively \cite{DSRC_overview}. 

Since DSRC was established as a standard for V2X communication in 2010,  it has been extensively integrated into numerous EV charging systems. For instance, in SWPT communication, the researchers in \cite{DSRCbasedSystem_Access} proposed a decentralized EV charging system (DEV-CC) that relies on DSRC to promote communication among EVs for charging resources coordination. Similarly, \cite{EV_congestion_pricing_charging_DSRC} designed a system that employs DSRC to offer positive feedback for drivers who opt for less congested routes and charging stations.  Another innovative approach was developed by researchers for power management in DSRC-based DEV-CC  \cite{powermanagement_DSRC}. The high data rate and low latency of DSRC, afforded by the 5.9 GHz frequency, make it particularly efficient in the DWPT scenario.  For example, \cite{s21} utilized DSRC to orchestrate information exchange between EVs and RSUs, establishing a dynamic EV charging system for authentication and payment.

While many proposed research and applications exist for EV charging systems over DSRC, the long-range communication issue remains a barrier. Due to the inherent spectral limitations of DSRC, the achievable effective communication range is confined to 1 kilometer. Consequently, for DSRC-based EV charging systems to maintain reliable communication, a higher density of RSUs is needed. This augmentation in RSU density subsequently results in significant cost escalations, posing a considerable challenge to the widespread implementation of DSRC-enabled EV charging systems.

\textbf{C-V2X}:
To enable communication among vehicles, pedestrians, and mobile end devices, 3GPP defines C-V2X to set up communications grounded on the existing cellular network \cite{Tutorial_5GNR_CV2X}. Hence, the communication among vehicles and end devices share the communication load or exchange information directly to achieve EV charging authentication, resource allocation, or grid management. These years, 3GPP also standardized new radio V2X (NR-V2X), which can enlarge its coverage, quality-of-service (QoS), and data rate.

Recently, C-V2X developed NR-V2X, which utilizes a frequency band that covers long-range and short-range communications where the frequency range is 450 MHz to 7.125 GHz and 28 to 52 GHz. Specifically, the short-range communications in 5G-V2X perform with millimeter-wave, increasing the throughput of the system with a higher data rate. With the enlarged effective communication range, the NR-V2X EV charging system coordinates the charging unit expediently. Besides, the dedicated communication equipment density is decreased by expanding the communication range on C-V2X EV charging communication systems.

However, C-V2X costs a lot in terms of maintenance and facility costs. In rural areas, the C-V2X equipment may consume much energy while the vehicular traffic density (VTD) is relatively low. As a result, the extra cost of EV charging communication systems yields fewer. Besides, the cellular communication system has spectrum limitations. With the increasing frequency bandwidth, the interference over the air impacts the reliability, which may cause the EV charging identification to be mislabeled by the errors.

\subsubsection{Other communication in V2X}
Except for DSRC and C-V2X, other standards have been performed in many other fields, like ZigBee and WiFi. Some of them are low-speed communication standards, which are not applicable in high mobility environments in V2X, while they afford ancillary communications in V2X. This subsection introduces these standards and summarizes the current application or research related to EV charging or similar fields.

\textbf{ZigBee}:  Developed by the Connectivity Standards Alliance (CSA) for low-cost, short-range applications, ZigBee is designed to facilitate smart home systems, internet of things (IoT), and industrial information interfaces \cite{ZigBee_security}. ZigBee provides three different network topologies for wireless communication, including star, tree, and mesh, which allows for a more flexible communication system, leading to wide utilization in industrial IoT \cite{ZigBee_Mesh}.

Additionally, ZigBee is a supplementary communication method for EV charging due to its low power consumption and complexity. For instance, in \cite{ZigBee_WPT_app}, ZigBee was tested to locate vehicle positions within a WPT system. Furthermore, a method to reduce the workload on cellular networks using ZigBee and WiFi was proposed in \cite{ZigBee_WiFi_V2X_app}.

However, ZigBee's low complexity in terms of memory size presents a security vulnerability. It is susceptible to attacks like reconnaissance, device manipulation, DoS, and network control \cite{ZigBee_security}. For example, a passive inference attack on ZigBee was proposed in \cite{Zigbee_zleaks}, which could discern events and devices in smart home applications. Another study identified threats posed by low-rate DoS attacks on indirect transmissions \cite{Zigbee_LDoS_indirectTrans}. Thereby, there are better choices than ZigBee to transmit private information in EV charging. 

\textbf{LoRa}: As a standard promoted by LoRa Alliance, LoRa has long-range wireless communication, low power consumption, reliability, low cost, and simple deployment. 

Due to the chirp spread spectrum and low-density parity-check code (LDPC) techniques that LoRa has, it can be performed well even in high interference environments \cite{CSS_LDPC_LoRa}. \cite{LoRa_Noisyenv} found that LoRa can detect the signal within a noisy environment. When EV charging happens in an urban environment, many kinds of interference are caused by mobile phone users, vehicle monitors, and other electrical devices, so the ability to perform weak signal detection is needed. Moreover, \cite{Infocom_weakLoRa_detection} proposed using stochastic resonance to enhance the SNR of weak LoRa signals and combine with the chirp characteristics of LoRa to distinguish the weak signal.

By its capacity to resist disturbance, LoRa can transmit highly long distances, which makes LoRa applicable in long-range communication requirement scenarios. \cite{LoRa_forest_urban_vehicle} investigate the LoRa's electromagnetic propagation in forest, urban, and vehicle communication environments. They found that communication can still work even if the distance has reached 200 meters in high-density building environments. For example, in smart grid management, the LoRa manages the aggregator and EV communication \cite{LoRa_aggregator_V2G}. In addition to the feature above, LoRa provides a security frame for communication. In \cite{LoRa_Edgecomp_secure}, they built up an EV charging smart grid system with decentralized LoRa as a communication protocol to ensure secure information transmission. 

Besides, in EV charging, especially static, one needs to gather information from edge devices like charging points and billing systems, and it requires the edge devices to work for a long time with low power consumption and reliable transmission. Therefore, security and the lifetime of devices need to be considered. Fortunately, as with ZigBee, LoRa has a long lifetime and low cost, which means LoRa can be easily employed widely on infrastructures and works for a long time without frequently updating. To prolong the lifetime of LoRa, \cite{LoRa_lifetime_linklayer} concentrated on the link layer protocol and multiple transmission parameter adjustments to extend the LoRa network lifetime. In the work proposed by \cite{LoRa_lifetime_AdapLoRa}, they extended the LoRa devices' lifetime by focusing on the resource allocation of LoRa by the quality of links for edge devices. Accordingly, in remote or noisy environments, EV charging devices rely on LoRa to render an effective communication strategy, which enables them to save costs and maintain QoS.

Beyond the performance of LoRa technology in noisy and remote static communication contexts, its ability to address challenges associated with mobility and latency in communication is equally essential. In a research conducted by \cite{LoRa_mobility_performance}, the effects of mobility on LoRa communications were undertaken.  The researchers found that the efficiency of LoRa diminished to a relatively lower level when subjected to increasing mobility. The results show QoS within an acceptable range if only the length of the transmitted information and the data rate at a lower level.  Consequently, these constraints reveal further exploration is needed to enhance the effectiveness of LoRa technology in dynamic communication scenarios.

\textbf{Bluetooth}:
As a widely used communication technology, Bluetooth is easily employed on mobile phones, computers, and public service facilities. Due to the convenience of deployment, Bluetooth has deployed on massive end devices so far \cite{blue_Tooth_D-V2X}. After the 4.0 versions, there are two types of Bluetooth technology: classic Bluetooth and Bluetooth Low Energy (BLE). The classic type has a higher speed with symmetrical topology, while BLE has a relatively low data rate, lower energy cost, and more topology structures \cite{Bluetooth_2019survey_multihop}.

Leveraging the extensive user base and the compatibility of Bluetooth, it is feasibly deployed to facilitate the implementation of small-scale Internet of Things (IoT) systems, as discussed in the survey by \cite{Bluetooth_survey_manyapp}. Similarly, Bluetooth is found in EV charging systems applications. The charging system explored in the study by \cite{Bluetooth_EV_authentication} is predicated on Bluetooth. It generates the shared secret between the OBU and the driver's smartphone. In contrast, a hashed value of this secret, including ID information in EV charging, will be delivered to the DSRC system. 


BLE, in version 5.0, introduces a more diversity of topological configurations for communications as compared to its symmetrical structure predecessors, offering point-to-point (P2P), centralized, broadcaster, and meshed structures \cite{Bluetooth_securing_survey}. These diverse structures allow BLE implementation with unique requirements in various EV charging communication scenarios. The centralized topology ensures communications security, while the broadcaster and meshed topologies escalate the communication range and robustness. Thus, BLE provides a list of structural options for communication network construction within the EV charging resource management framework, satisfied with a spectrum of demands.  Therefore, these advances in Bluetooth technology contribute significantly to its adaptability and efficiency in diverse communication environments, reinforcing its applicability in certain areas.

\textbf{WiFi}: 
WiFi has quickly become the dominant standard in wireless communication fields recently \cite{wifi_intro_defi,wifi_survey2016}. Especially the next generation of WiFi, WiFi 6/6E, which WiFi Alliance with standard IEEE 802.11ax designs, has a faster data rate, lower latency, and higher spectrum efficiency than WiFi 5 \cite{WiFI6_firstlook}.

Since there is no restriction on spectrum for WiFi \cite{wifi_V2X}, which means devices with different WiFi generations can communicate without extra escalation. However, WiFi provides high throughput with a 9.6 Gbps data rate in WiFi 6/6 E \cite{WiFi6_throughput}. In \cite{Mobisys_scatter_wifi}, the researchers designed a backscatter system over commercial WiFi devices and achieved a higher data rate than the limited throughput. Besides, the OFDMA in WiFi 6/6E divides the entire frequency band as resource units (RUs) to each end device. This measure helps to reduce communication overhead while the connectivity is enlarged \cite{WiFI6_firstlook, WiFi6_5G_compare}. Therefore, it is sufficient for WiFi to support massive EV charging communication in high-density areas.


In some research, the VANET helps to set up communication between vehicles and charging stations \cite{VPS_EV_charging}. Although the traditional VANET is based on DSRC or C-V2X, WiFi direct is also one of the potential wireless technologies for VANET \cite{VANET_wifidirect}. They introduced the use of WiFi directly to expand the capacity of C-V2X in VANET, especially in high VTD areas, like urban areas and centralized office places.

\subsection{Communication standard vs EV charging communication requirements}
In this subsection, we analyze the communication standards for both static and dynamic EV charging scenarios. Recognizing the unique requests of each scenario, we identify and include communication standards that may fulfill the respective requirements, thereby offering potential solutions for vehicle communication in each context.

\subsubsection{Communication for SWPT requirements}
SWPT requires more metrics on high-density connections, security, and reliability. Unlike a dynamic situation, to ensure traffic security and avoid collision, the density of vehicles is constrained and sparse in space. As for the static part, especially in parking lots, the design of these places is to contain vehicles as many as possible. Therefore, the utility rate of space is higher than on highways. Besides, in traffic-congested urban areas, the density of vehicles increases to relatively high levels as well. Because of these factors, a massive connection is required in these scenes, and the sufficient connectivity of LoRa, ZigBee, WiFi, and BLE can take this responsibility. However, due to the energy efficiency, security, and data rate requirements, LoRa and BLE are more satisfied with this trade-off in static scenes. Apart from the communication standards mentioned above, the RFID and Ethernet are also conducive to building communication in static EV charging \cite{RFID_EV_charging,ethernet_EV_charging}. Because of the security and unlimited spectrum of wired communication, the message delivered via wired medium may be a better choice.

\subsubsection{Communication for DWPT requirements}
Different from SWPT, DWPT requirement focuses more on its mobility and latency. According to the analysis in subsection \ref{sec4_1}, the DSRC, C-V2X, and WiFi are more suitable. All three communication standards afford sufficient data rates and lower latency. Especially due to the more extended communication range of DSRC and C-V2X, they can cover more traffic areas quickly with the existing communication facilities. So, both C-V2X and DSRC are the primary communication standards for EV charging systems to be the proper choice for DWPT communications. 

But when it comes to urban areas, the traffic congestion road is another dynamic scenario while vehicles move slowly. It needs real-time communication and also robust communication in a short range. Therefore, WiFi can be a good choice. As the analysis we mentioned in subsection \ref{sec4_1}, it provides a high data rate and low latency by WiFi 6/6E. At the same time, it is suitable and reliable for short-range mobile communications \cite{massive_connection_required}. However, considering the relatively higher cost of WiFi devices, these WiFi EV charging systems can only be allocated in small-scale areas prone to congestion, thus requiring proper scheduling \cite{lihao}.

\section{Communication to grid}\label{sec5}

\subsection{WPT impact on grid}
With the growing number of EVs, the charging demand is increasing. This creates enormous pressure on the grid, especially during charging peak times. The grid is designed to be available, stable, and reliable. The SWPT and DWPT can impact the grid differently due to their specific charging scenarios. Integrating EV charging into the grid system is challenging \cite{garwa2019impact}. 

Like wired charging, SWPT is usually designated to take place in a charging station or at home. When charging at home, the charging time is likely at night. While charging in the charging stations, the time is more likely to be during the day. The extensive parking lots with charging infrastructures can make the charging load heavier during the day. The power source for SWPT can be directly from the grid or indirectly from the energy storage in the charging station \cite{yan2018optimized}. With the help of energy storage, the charging peak to the grid can be reduced. Also, energy storage allows renewable energy sources such as solar panels to be used. In \cite{zhang2018optimal}, a charging station with dual charging modes is considered. Low-power AC charging has a longer charging duration at a cheaper price. High-power DC charging is faster but comes at a higher cost. The future wireless charging stations will also provide different charging services.  

For DWPT, the primary pad is deployed along the road, and there is usually no energy storage for the primary charging pad. The charging infrastructure is connected to the grid directly. The traffic and the charging demands of DWPT are unpredictable. When a large number of vehicles use DWPT on the road, it may cause significant challenges for the grid. In \cite{debnath2018grid}, the authors show that DWPT can significantly impact the grid voltage. The variations in grid voltage can reduce power transfer efficiency. The authors in \cite{zeng2021grid} analyzed 24-hour charging load and grid voltage profiles. The grid voltage changes oppositely from the charging load. 

The impact of WPT on the grid requires the grid to be stable to support WPT applications. Smart grid controls can be applied to achieve a robust, stable grid. The grid operators need to know the power demand and smart control the power allocations. Different communication techniques can be used to collect the power usage data and help the grid operators to predict the power demand. In \cite{mehar2023wireless}, the smart grid is mentioned to make EV charging easy. The smart grid uses two-way communication and distributed smart devices to monitor the grid status and control the power flow. The authors in \cite{wang2019extended} presented a WPT system in the smart grid environment. The WPT system is applied with adaptive resonant frequency tracking, and the grid is employed with automatic load optimization and output voltage regulation. With intelligent control, the power transmission efficiency and robustness can be improved. 

\subsection{Vehicle-to-Grid (V2G)}
Another method to stabilize the grid is V2G. In contrast to charging, V2G transfers energy from the EV battery to the grid. It is achieved by bi-directional charging. The power can be transferred from the power grid to the EV battery and vice versa. The power electronics of EVs are more complicated due to the bi-directional power transfer. This makes the EV battery work like a mobile power energy. Since EVs are geographically distributed, they can be distributed for battery energy storage. The EVs can perform charging while the grid is in off-peak conditions and perform V2G while the grid is in peak demand. There are multiple advantages of V2G, including assuaging the rapid spikes, stabilizing voltage, shifting peak load, and regularizing frequency \cite{iqbal2020v2g}. When the charging pattern is known, V2G can be applied autonomously to satisfy scheduled charging requests \cite{ota2011autonomous}. The EVs are connected to the grid to perform G2V or V2G when needed. A regional energy management system can control this. Previous research focused on the V2G under wired connection scenarios. Most recently, due to the advancement of WPT, V2G design is proposed in wireless connection scenarios \cite{8723851, yang2017design, ahmad2019bidirectional}. The electronics design is similar to the wired V2G. The energy efficiency of V2G is comparable to G2V in WPT. In \cite{azad2019bidirectional}, the V2G is applied in DWPT scenarios. Extra control is needed to achieve that.

Repeatedly charging/discharging EV batteries may shorten the battery life. So, economic costs should be considered when scheduling V2G activities. In \cite{zhang2021electric}, the authors proposed a dynamic V2G scheduling framework by considering the EVs and the grid together. The variance of grid power is mitigated with relatively low economic costs. The researchers in \cite{almehizia2018investigation} investigated the economic viability of V2G. It is shown that V2G can achieve significant positive economic effects in peak demand time. Incentives such as free night charging can be utilized to encourage EV owners to participate in the V2G program. 

To achieve highly efficient V2G, the communication between the EV and grid and the grid communication is significant. In \cite{huang2013interaction}, the researchers investigated the interaction between the EV and the grid with wireless links. The high-level V2G system structures are designed. A review of communication standards of V2G is presented in \cite{vadi2019review}. The communication pin in the charging socket is used for communication via EVSE. While in \cite{saltanovs2017analysis}, wireless communication is analyzed in V2G applications. IEEE 802.11 WiFi-based wireless communication is regarded as the potential communication standard. The communication delay is applicable in SWPT scenarios.  

\subsection{Grid communication}
The communication between EV and the grid can be wireless, as mentioned in Section \ref{sec4}. Grid communication can also be realized via wireless networks or power lines. The PLC uses the existing infrastructure, the power line, as the transmission media. It is cost-effective without extra deployment. The wired connection makes it more secure than wireless networks to deliver the charging demand information. However, PLC can be easily affected by noise from other electrical appliances. And the small bandwidth restricts the transmission capacity. In \cite{ma2013smart}, the smart grid communication is reviewed. The electrical grid, communication infrastructure, and control center are integrated to make the grid smart. The gird information is collected and transferred via the communication infrastructure to the control center. The control center processes the messages, and the control commands are sent back. Different communication protocols can be used to satisfy diverse communication needs. Wide area networks (WAN), such as cellular and WiMAX, can communicate multiple local grids. The local area network (LAN), such as WiFi, can be employed within the local grid.

To mitigate the overall delay for DWPT, the delay of grid communication should also be considered. The communication delay of the power grid is analyzed in \cite{muyizere2022effects}.  The delays come from several stages, including transmission from the measurement unit to the control center, computing, and phase synchronization through the Global Positioning System (GPS). The sudden delay can be reduced by a robust delay design. At the same time, the inherent delay needs to be addressed by advanced communication protocols.

\section{Security and privacy }\label{sec6}
In this section, we will discuss the principal security challenges faced by WPT systems and outline existing state-of-the-art security solutions by analyzing different types of attacks and their potential consequences, as well as assessing the effectiveness and applicability of current defense methods.

\subsection{Security vulnerabilities}
As the adoption of EVs and DWPT systems keeps growing, security solutions must scale efficiently to cater to an expanding user base and infrastructure components. WPT systems encompass financial transactions, critical control message exchanges over wireless networks, and real-time operations amid dynamically changing environments and unattended devices.  Due to the broadcast nature, wireless communications among parties such as EV, transmitter pads, the power transmitter, charging station (CS), and RSU should provide efficient security measurements in a real-time manner while protecting the private information of EV drivers. For example, the transmitter pads should only switch on during the charging period of authorized EVs to save energy and prevent energy theft by unauthorized EVs \cite{shj}.

The wireless mode of communication in DWPT systems opens avenues for various attacks, such as eavesdropping, man-in-the-middle, and replay attacks, making establishing secure wireless communications a paramount challenge. 
Safeguarding user privacy, encompassing location and financial information while facilitating seamless charging operations, is significantly challenging. 
As listed in table \ref{tbsec}, the critical security aspects required for WPT systems protection should include confidentiality, authentication, integrity, non-repudiation, message freshness, availability, backward secrecy, and forward secrecy.
For example, ensuring the integrity of data and financial transactions within the system to retain their accuracy, completeness, and resistance to unauthorized alterations is a notable challenge, especially within the wireless and dynamic milieu of DWPT systems.

The unique nature of the DWPT system makes designing effective security defense methods more challenging. 
The mobility of EVs and the real-time demands of charging necessitate security measures that operate in near real-time to thwart and mitigate attacks, requiring the development of efficient, low-latency security protocols. 
Precise and robust mechanisms for authentication and authorization are imperative to deter impersonation and unauthorized access to charging services, amplified by the dynamic and wireless nature of DWPT systems. Moreover, implementing robust security measures while adhering to resource constraints like processing power and bandwidth, particularly in high-speed and real-time operational scenarios, poses a substantial challenge. Each of these challenges demands a meticulous and integrated approach to ensure the security and reliability of WPT systems, promoting their widespread adoption and subsequent success in the evolving landscape of electric vehicle charging.


\begin{table}
\centering
\caption{Security aspects in EV system. \label{tbsec}}
\begin{tabular}{| c |c|} \hline 
Security Aspect & Description \\
\hline
Confidentiality & Protects sensitive data during vehicle to\\ 
&vehicle or vehicle-to-infrastructure\\ &communications, allowing certain public data. \\
\hline
Authentication & Verifies system participants to prevent \\
&unauthorized access and fraudulent\\
&data transmission. \\
\hline
Integrity & Ensures data remains unchanged during \\

&transmission, guarding against manipulation\\
& and distortions. \\
\hline
Non-Repudiation & Ensures transmitted messages can't be\\
&denied by EVs in critical situations. \\
\hline
Message Freshness & Guarantees real-time delivery of emergency\\
&alerts and signals, thwarting replay \\
&and time-based threats. \\
\hline
Availability & Maintains system access to legitimate users,\\
&even during network failures or DoS attacks. \\
\hline
Backward Secrecy & Prevents newly joined EVs from accessing\\
&messages exchanged before \\
&their presence on the network. \\
\hline
Forward Secrecy & Preserves message secrecy as EV roams\\ 
&across multiple charging stations, \\
& even post-interaction. \\
\hline

\end{tabular}

\end{table}

Current wireless communication technologies associated with DWPT systems are still under development, unveiling several security vulnerabilities. Interactions between EVs and charging entities have brought to light many attacks, imperiling the confidentiality, authentication, integrity, and availability of the EV charging infrastructure. Noteworthy among these threats are impersonation attacks, replay attacks, DoS attacks, jamming, eavesdropping attacks, etc \cite{9693972}, which have the potential to precipitate data breaches, severely disrupt the EV charging process, and consequently result in financial losses and personal injuries.


 \subsection{Classic attacks and defense methods}
To ensure the safe operation of WPT systems, a comprehensive security assessment is imperative, along with the exploration of effective security solutions. Several countermeasures have been proposed to address the above attacks, including data authentication, encryption, position verification, and digital certificates. 
Adopting lightweight cryptographic protocols, utilizing privacy-preserving technologies such as pseudonym systems, and implementing mutual authentication and session key protocols can further safeguard the communication process. Additionally, some researchers have introduced authorization protocols based on the physical layer and secure management strategies leveraging natural energy resources to reduce operational costs and enhance system security.

Impersonation attacks emerge as a particularly concerning threat vector. They manifest in two principal forms: vehicle impersonation and server impersonation. 
In vehicle impersonation attacks, attackers fabricate identities to transmit messages to servers, masquerading as legitimate vehicles. 
In server impersonation attacks, a malicious server dispatches falsified data to vehicles, pretending to be legitimate servers. 
To tackle these issues, a fast authentication protocol for wireless charging between EVs and charging pads (CPs) was proposed in \cite{s6}, leveraging symmetric keys and the spatiotemporal location of EVs. Additionally, a physical-layer-assisted security scheme was introduced in \cite{s7}, employing a hierarchical authentication approach among various parties, such as banks, CS, RSUs, CPs, and EVs.

Replay attacks encompass the unauthorized reiteration of previously sent communications to gain illicit access to network services and resources. 
Replay attacks were addressed in \cite{s11} through a fast authentication for dynamic EV charging with fast signing and verification. This protocol also protects against man-in-ihe-middle (MITM) and impersonation attacks, ensuring a robust defense against unauthorized access and data tampering.

Injection attacks involve the transmission of legitimate messages to gain control, with the potential to transmit malicious messages after that. 
A private blockchain was designed in \cite{s9} to support privacy-preserving dynamic charging coordination, authentication, and billing, which could resist false information injection, advertising fraudulent energy services, and replacing with poor-quality energy.

Double-spending denotes a scenario where a user repeatedly utilizes the same token at a charging point during authentication.
Double-spending was addressed in \cite{s18} by proposing an authentication protocol that resists such issues.

Privileged insider attack pertains to trusted network users exploiting acquired secret credentials, underscoring the need for robust protocols to thwart insider credential theft. In \cite{sinsider}, researchers proposed a secure and robust blockchain-based electric vehicle charging system using the ECC cryptosystem and lightweight one-way hash functions, which can withstand significant attacks, including Privileged-insider attacks.

Further amplifying the security conundrum are DoS attacks, where adversaries overwhelm servers to deny access to legitimate users—a menace that amplifies risks during emergencies. The IEEE 802.11p standard, which utilizes the elliptic curve digital signature algorithm (ECDSA) for authentication in-vehicle networks, was noted in \cite{s5} for its susceptibility to denial-of-service attacks due to the time-consuming nature of signing and verifying a signature.

Jamming attacks endeavor to disrupt wireless connections between EVs and entities like charging points and RSUs, thereby threatening wireless communication. 
In \cite{sjam}, the authors introduced key agreement protocols that do not require security infrastructure support and can generate cheap but random enough secret keys. The introduced jamming-resilient protocol provides robust throughput even in the presence of intelligent jammers.

An eavesdropping attack sees attackers intercepting confidential data from network communications without participation, while message tampering involves the alteration of message structures to influence recipient decisions, potentially immobilizing the system. Eavesdropping attacks were highlighted as a significant threat, especially when authentication messages are converted into bits at the physical layer. This vulnerability, discussed in \cite{s20}, allows eavesdroppers to overhear and intercept signals, exposing the system to various security threats.

Message tampering, where attackers modify message structures to influence recipient decisions and potentially halt the system, was tackled in \cite{s13} by proposing an authentication scheme that combines different cryptosystems, such as hashing and exclusive-or operations, to improve the security of dynamic charging systems while preserving the privacy of the drivers.

Additional attack vectors include guessing attacks, where attackers deduce user identity or biometrics from intercepted messages or lost/stolen OBUs or smart cards. A known critical attack involves linking previously generated authenticated keys to extract pivotal information. A session linking attack entails connecting randomly generated sessions to disclose credentials through rudimentary algorithms, revealing sensitive information.

\subsection{Realtime defense methods for comprehensive attacks}

The wireless links between the EV and the charging infrastructure allow attackers to intercept messages transmitted by the EV or power transfer unit, potentially culminating in fraudulent delay or redirection of these messages. The inherent mobility of EVs mandates that security measures in DWPT systems operate in near real-time, given the limited time window available for both communication and security protocols. This difficulty is further exacerbated by the mobility of the involved entities, demanding a meticulous and integrated approach to fortify the security and reliability of DWPT systems. Such fortified security is pivotal for promoting widespread adoption and ensuring the subsequent success of EV charging systems in the evolving electric vehicle charging landscape.

In our previous work \cite{shj}, we proposed authentication and physical layer security schemes to improve secure communications between the EV and charging infrastructure in DWPT systems. In particular, a double-encryption with the signature (DoES) scheme is proposed for session key exchange between EV and charging station, which provides data authenticity and integrity. To enable low-latency authentication between EV and power transmitter in DWPT systems, a sign-encrypt-message (SEM) authentication code scheme is designed leveraging symmetric keys for dynamic charging, which ensures privacy and resistance to tampering attacks. The artificial noise-based physical layer security (ANbased PLS) scheme is also proposed at the physical layer to degrade the wiretapped signal quality of multiple eavesdroppers operating in non-colluding and colluding cases. Closed-form expressions for the secrecy outage probability (SOP) and intercept probability (IP) of the considered system with the non-colluding case are derived to show that the proposed AN-based PLS scheme provides lower SOP and IP than the conventional ones without AN. The distance between eavesdroppers and the power transmitter also affects the system SOP and IP in non-colluding and colluding cases. Moreover, the EV using the DoES scheme takes 52 ms to obtain session keys from the charging station, while it only uses 8.23 ms with the SEM scheme to authenticate with a power transmitter for the charging process. The DoES scheme was proven to resist various attacks in DWPT systems, such as man-in-the-middle attacks, replay attacks, forward secrecy attacks, insider attacks, and eavesdropping threats. At the same time, the SEM was shown to guarantee low latency, privacy requirements, and resistance to tampering attacks. The proposed AN-based PLS scheme could decrease the signal quality received at non-colluding eavesdroppers with low SOP and IP, which protected the DWPT system from eavesdropping attacks. 


Introducing blockchain technology in DWPT systems can help address the concerns above. This technology, as explored in \cite{s9}, \cite{s14}, \cite{s19}, and \cite{s21}, employs distributed consensus processes and cryptography to achieve security and privacy protection. Although blockchain-based approaches have shown promise in achieving privacy and security, they involve third-party inclusion, creating complex exchanged information and delays between the EV and CS. Despite these advancements, the security of wireless charging systems at the physical layer has not been extensively studied in previous works, leaving room for further exploration and improvement in securing DWPT systems against the diverse range of attacks discussed.

\section{Trend, challenges, future directions}
\subsection{Trend}
According to the International Energy Agency (IEA) \cite{iea_ev_global_report}, the number of electric vehicles has grown exponentially during the past decade. The number of EVs sold in 2010 is negligible, while in 2022, it is over 26 million. Based on the trend, more and more EVs will travel on the road. This will bring a significant charging demand. More charging stations and fast, novel technologies are needed to satisfy the charging demand. The charging power of the SWPT and DWPT will also be defined to satisfy different use cases. For users with a fast charging demand, high-power SWPT can provide fast charging speed. Or DWPT with optimal routes can charge EVs with more time. For SWPT at home, slower charging with less infrastructure cost can be deployed for overnight charging.


The charging and communication standards for wired charging are defined in SAE J1772, CCS, and CHAdeMO. EVs from different brands can be charged in the same station using the same connector. However, multiple sockets need to be installed on the EVs to charge under different standards. For wireless charging, there are no unified standards now. The charging and communication methods are still under research or in the pilot period. However, due to the advantages of wireless charging, such as safety and friendly usage, wireless charging demands will grow fast. The charging and communication standards will finalized to reduce the infrastructure costs. The 
universe wireless charging standard will allow vehicles to be used without charging limitations. Only one wireless charging interface is required. Also, the novel DWPT method to charge EVs in motion will accelerate road infrastructure improvement. In \cite{yokoi2023current}, the author analyzed the electric road system (ERS) for DWPT and envisioned the future implementation. When more primary charging pads are deployed on the road, DWPT can become the primary method for EV charging. This can reduce the EV battery weight while extending the driving range. 

\subsection{Challenges}
While the WPT is a promising technology that provides better safety and reduces driving range anxiety, it still faces many challenges. The charging efficiency of WPT is still lower than wired charging. According to \cite{bi2016review}, the wired charging can achieve above $90\%$ energy efficiency. At the same time, the charging efficiency is around $85\%$ to $90\%$ for SWPT and $72\%$ to $83\%$ for DWPT. Extra engineering and research are needed to improve energy efficiency. 

Also, extra charging pads need to be deployed on the ground and under the vehicle to fulfill WPT. It will bring infrastructure costs to the operator. The charging pad attached to the vehicle will also increase the vehicle's weight and reduce the driving range in some way. For DWPT, constructing roads with primary charging pads installed is a big project that will bring huge costs. Also, the scheduling of the EV in DWPT scenarios involves the integration of grid preparation, EV status sensing and control, and communication. Multiple systems need to work together properly to build an efficient, robust, and secure charging system. 

The DWPT has higher requirements for the grid due to the high mobility and low latency of EVs. The grid needs to prepare for the instant peak times caused by multiple EVs charging in DWPT. This will cause significant pressure on the grid. 

The existing communication standards are designed to fulfill EV wireless charging, which may not fit in the future WPT. For SWPT, LoRa, ZigBee, WiFi, and BLE are considered. However, Lora is designed to provide low-power transmission, which caused the data rate to be less than 27 kbps. The maximum number of connections for the BLE host is 20, which only supports 20 EVs connected. Zigbee and WiFi use the same 2.4 GHz carrier frequency, which may have interference problems. For DWPT, the DSRC and C-V2X are considered as the potential standards. DSRC is based on IEEE 802.11p to provide low-latency data transmission. It supports the high mobility of EVs in motion. However, the communication range of DSRC is around 400 m and up to 1000 m \cite{DSRC_intro}. The short communication range requires more DSRC RSUs to be installed. While C-V2X employs the existing cellular infrastructure to reduce infrastructure costs, C-V2X has higher synchronous requirements and symbol duration. The symbol duration is 8 $\mu s$ for DSRC, while 66.67 $\mu s$ for C-V2X. This causes the C-V2X high latency in transmission. According to \cite{maglogiannis2021experimental}, the average end-to-end latency is 5.5 ms for DSRC while 37.84 ms for C-V2X PC 5 at 97$\%$ packet transmission.  DWPT requires ultra-low latency and higher reliability. The existing DSRC and C-V2X do not fully support these requirements, and a new communication standard designed for DWPT is needed.

\subsection{Future directions}
Some new technologies are emerging to speed up the adoption of WPT for EVs. Simultaneous wireless power and data transmission (SWPDT) is one of the promising techniques to provide both charging and communication services for WPT. Also, 5G-based C-V2X can reduce the communication latency for DWPT compared with 4G-based C-V2X. 

SWPDT aims to reduce the costs of communication infrastructures by sharing the existing power coils for wireless charging. The power and data are transmitted simultaneously from the primary to the secondary pad. Using two data carrier frequencies, the authors in \cite{qian2019full} achieve full-duplex communications between the EVSE and EVs. A 64 kbps data rate is sufficient for WPT communication. In \cite{casaucao2023simultaneous}, a review of SWPDT for EV charging is given. The SWPDT is classified based on different criteria: 1) number of links and signal carriers, 2) data communication type, 3) signal multiplexing technique, 4) modulation of data signal, 5) data injection/extraction, and 6) compensation system. This area remains for further research and experimental validations. Specifically,  SWPDT in dynamic charging and security analysis of SWPDT systems have not been studied. 

With the development of cellular networks, C-V2X is evolving from 4G long-term evolution V2X (LTE-V2X) to 5G NR-V2X. The multiple access technique is changing from SC-FDMA to OFDMA \cite{sehla2022resource}. New numerologies and multiplexing of the physical sidelink control channel (PSCCH) and physical sidelink shared channel (PSSCH) favor reducing the communication latency. The latency as short as 3ms is supported by 5G NR-based V2X in specification Rel-15 \cite{shin2023vehicle}. Also, the joint use of DSRC and C-V2X in the 5.9 GHz ITS band is proposed for V2X communications \cite{ansari2021joint}. The DSRC and C-V2X can be combined as a hybrid V2X system to provide user service.

\section{Conclusions}

This review examines the communication requirements of static and dynamic wireless power transfer for electric vehicles. Existing communication standards, including cellular, WiFi, and DSRC, can meet specific communication requirements for EV WPT. The communication with the grid to facilitate EV WPT is also analyzed. Next, we discuss the security and privacy issues associated with EV WPT communications. Finally, the challenges and future trends of communication for EV WPT are analyzed. This can work as a guideline for future research on EV communication integration.

\vfill\pagebreak

\section{Acknowledgment}
H. Sun would like to thank the support from UGA E-Mobility Initiative and NSF-2236449. 
\bibliographystyle{iet}
\bibliography{refs.bib}

\end{document}